# Testing cosmic homogeneity and isotropy using galaxy correlations

Michael J. Longo[1]

Department of Physics, University of Michigan, Ann Arbor, MI 48109-1040

Despite its fundamental importance in cosmology, there have been very few straightforward tests of the cosmological principle. Such tests are especially timely because of the hemispherical asymmetry in the cosmic microwave background recently observed by the Planck collaboration. Most tests to date looked at the redshift dependence of cosmological parameters. These are subject to large systematic effects that require modeling and bias corrections. Unlike previous tests, the tests described here compare galaxy distributions in equal volumes at the same redshift *z*. This allows a straight-forward test and *z*-dependent biases are not a problem. Using ~$10^6$ galaxies from the SDSS DR7 survey, I show that regions of space separated by ~2 Gpc have the same average galaxy correlation radii, amplitudes, and number density to within approx. 5%, which is consistent with standard model expectations.

## 1. INTRODUCTION

One of the pillars of modern cosmology is the cosmological principle that states that, on sufficiently large scales, the universe is homogeneous and isotropic. To date there have been surprisingly few tests of this principle. Most of these have looked for a redshift (radial) dependence of important cosmological parameters or have used galaxy density projections on the sky. The most reliable test is the isotropy of the cosmic microwave background (CMB). However, there are many indications that the CMB is not, in fact, completely isotropic. There are dozens of publications that suggest power asymmetries and unlikely alignments of low multipoles (see, for example, [1, 2, 3, 4, 5]). There are also indications of parity-violating asymmetries [6, 7]. Mariano and Perivolaropoulos [8] discuss several pieces of evidence for its violation: the apparent alignment of the WMAP asymmetry axes with those found in large-scale velocity flows [9], the fine structure constant dipole [10], and a dark energy dipole [11]. Each of these deviations from isotropy is between $2\sigma$ and $4\sigma$.

The WMAP asymmetries have been criticized as being due to deficiencies in the WMAP subtraction of the microwave background from the disk of our Galaxy. (See, for example, Bennett *et al*. [12].) Recently, however, more precise Planck satellite results have shown what appears to be a significant hemispherical asymmetry whose axis is almost 90° from the Galactic axis [13]. The other anomalies in the WMAP data were also seen in the Planck data, including a parity-violating asymmetry. The agreement between the two independent experiments effectively rules out the possibility that their origin lays in systematic artifacts present in either data set. The Planck/WMAP results therefore suggest a small violation of isotropy. This tension with the cosmological principle is the subject of much discussion (see, for example, [1, 14]), which makes other tests of cosmic homogeneity that use probes other than the CMB of particular interest.

---
[1] mlongo@umich.edu

One expectation of homogeneity is that the length scales and amplitudes associated with galaxy distributions are constant everywhere except for a slow variation with look-back time (or redshift). Direct tests of homogeneity that look at the matter distribution are difficult. Unlike the CMB, the galaxy counts only include visible matter, not dark matter. At the observational level, they do not cover a large part of the sky due to the obscuration by dust near the Galactic plane. Surveys of a large portion of the northern Galactic hemisphere have become available only recently, and extensive surveys of the southern hemisphere are still not available.

Cluster sizes and counts have been an important tool for understanding dark energy and other questions in cosmology. Most past studies of homogeneity using galaxy correlations have studied only the angle-averaged variation of galaxy distributions with redshift [15, 16, 17]. These suffer from the difficulty of accounting for the natural evolution with look-back time, as well as the changing composition of the observed galaxies with increasing redshift as the galaxies become fainter and redder, and other redshift-dependent distortions. These biases must be understood in order to study the evolution of the correlations with increasing $z$ [18, 19]. In addition, most of these studies are limited to distances ≲300 Mpc, and most use only 2D separations based on angular coordinates.

Techniques that require explicit cluster finding involve small samples (see, for example, [20]), and are also subject to serious redshift-dependent biases [21]. Galaxy counts have been used to look for mass concentrations or voids out to redshifts ~0.1, corresponding to distances ~300 Mpc [22], while the largest observed structure, the Sloan Great Wall, is centered near 234 Mpc [23].

This analysis avoids these problems by using equal-volume regions of the sky that are at the same redshift but widely separated in distance. It makes efficient use of all galaxies with spectra in the Sloan Digital Sky Survey Data Release 7 [24], and extends out to $z \approx 0.4$. It shows that the average size of galaxy clusters and clustering amplitudes can be measured to an accuracy ~5% to test for homogeneity over distance scales ranging from 200 to 2400 Mpc/$h$, where $h$ is a scale factor ≈0.7.

In this analysis, clusters are not identified explicitly. Instead, 3-D galaxy separation correlation lengths and amplitudes are used as proxies for cluster sizes and spatial densities. For sim-plicity, I refer to these statistical correlations as "clusters". All comparisons are made between equal volumes at the same redshift.

## 2. THE SDSS SAMPLE

The SDSS DR7 database [24] contains ~800,000 galaxies with spectra for $z \lesssim 1$. The sky coverage of the spectroscopic data is almost complete for right ascensions ($\alpha$) between 110° and 240° and declinations ($\delta$) between –5° and 60°. For the hemisphere toward $\alpha = 0°$, only narrow bands in $\delta$ near –10°, 0°, and +10° are covered. All objects with spectra that were classified as "galaxies" in the SDSS DR7 database were used in this analysis. The spectroscopic redshifts have redshift uncertainties $\lesssim 0.06\%$; this allows an accurate determination of the 3-dimensional separation of galaxy pairs that was essential in the analysis.

## 3. THE ANALYSIS

The galaxies were binned in redshift, and in right ascension, and declination ranges of equal volume. The 3-dimensional comoving distance between every pair of galaxies in each volume was calculated. For an $\Omega = 1$ universe the comoving distance between two closely spaced objects is given by the same expression as in Euclidean space. In Cartesian coordinates, in terms of right ascension, declination, and comoving line-of-sight distance $r(z)$, the coordinates are

$$x' = r_p \cos\alpha, \quad y' = r_p \sin\alpha, \quad z' = r\sin\delta \quad \text{where} \quad r_p = r\cos\delta \qquad (1)$$

This gives a right-handed coordinate system with the $x'$-axis along $\alpha = 0°$, the $y'$-axis along $\alpha = 90°$, and the $z'$-axis along $\delta = 90°$. The connection between the redshift and the comoving line-of-sight distance $r(z)$ is determined by the cosmological model chosen. Since $r$ is just a scale factor in calculating the galaxy separations, the choice of a model is not critical to this analysis. I use a flat cold dark matter (CDM) model with $\Omega_m=0.3$, $\Omega_\Lambda=0.7$. The 3D comoving separation $d$ between two nearby objects is then

$$d = \sqrt{\delta x'^2 + \delta y'^2 + \delta z'^2} \qquad (2)$$

where $\delta x'$, $\delta y'$, $\delta z'$ are the finite differences along the $x'$, $y'$, $z'$ axes respectively.

For each ($\alpha, \delta, z$) bin, the 3-dimensional separations were calculated and binned into 125 bins for separations between 0 and 0.010 in dimensionless units (approx. 0 to 30 Mpc). The separations were then calculated with the galaxies' right ascensions, declinations, and redshifts in each ($\alpha, \delta, z$) bin randomly scrambled among the galaxies. To reduce the statistical fluctuations in the scrambling process, 10 scrambling runs were averaged. As a check on this procedure, as many as 100 scrambled runs were tried; this caused no significant change in the resulting parameters. Figure 1 compares the ratio $R$ of the not-scrambled ($NS$) and scrambled ($S$) distributions for the number of pairs in each bin *vs.* separation $d$ for typical $z$ ranges. The peaking at small separations is a clear signal for galaxy clustering on a length scale ~0.003 in dimensionless units (~9 Mpc/$h$). The width of the peak is a measure of the average radius of clusters in the volume studied, and the amplitude of the peak or its integral is a measure of the clustering strength.

Occasionally in the SDSS data, large galaxies appear more than once with different IDs when different points in the same galaxy are chosen as centers. Therefore pairs with separations <1x10$^{-5}$ were excluded in order to remove possible duplicates. (This is several times the radius of a typical large galaxy.) The ratios were normalized to an average of 1.0 in the separation range 0.0064 to 0.010. The correlation function used here is

$$\xi(\alpha,\delta,z) = \frac{NS-S}{S} = R-1 \qquad (3)$$

where $R$ is the *NS/S* ratio. The solid curves in Fig. 1 are fits to an exponential, $R-1 = a_0 e^{-d/a_1}$, where $a_0$ is the amplitude of the exponential, and $a_1$ is a measure of the width or radius of the distribution. The integral of the exponential is $a_0 a_1$. A least-squares fit to the exponential using bins 2 through 60, corresponding to $0.000125 < d < 0.0048$, was made. The first bin was not used in order to avoid systematic effects due to the finite resolution of the SDSS camera (*e.g.*, "fiber collisions"). The exponential fit, though simple and convenient, is not physically motivated, and sometimes gave poor fits or failed completely. This is apparent at small separations in Fig. 1. Therefore, a model-independent, numerical measure of the rms width of the peak $r_{sum}$ was determined from the contents of the bins for separations $0.000125 < d < 0.0048$. As a measure of the clustering strength, $I_{sum}$ is defined as a numerical integral or sum of the bin contents. Generally the radii and integrals from the exponential fit tracked those from the histogram bins very well. However, those from the contents of the bins typically had somewhat smaller uncertainties and were more robust when the bin occupancy was small. In the following, only the radii and integrals from the sum over bins will be given.

The uncertainties in the fitted radii and integrals were estimated in several ways. Statistical errors could be estimated from the contents of the separation bins with the correlations between the points taken into account. A jackknife resampling method using 10 non-overlapping subsamples, each with 10% of the data, was also used to estimate uncertainties. The resampling runs were done with the same programs as the normal runs with the same ($\alpha$, $\delta$, $z$) bins. The resampling method typically gave uncertainties ~50% larger than the statistical ones. In the following, to be conservative, the resampling uncertainties are given.

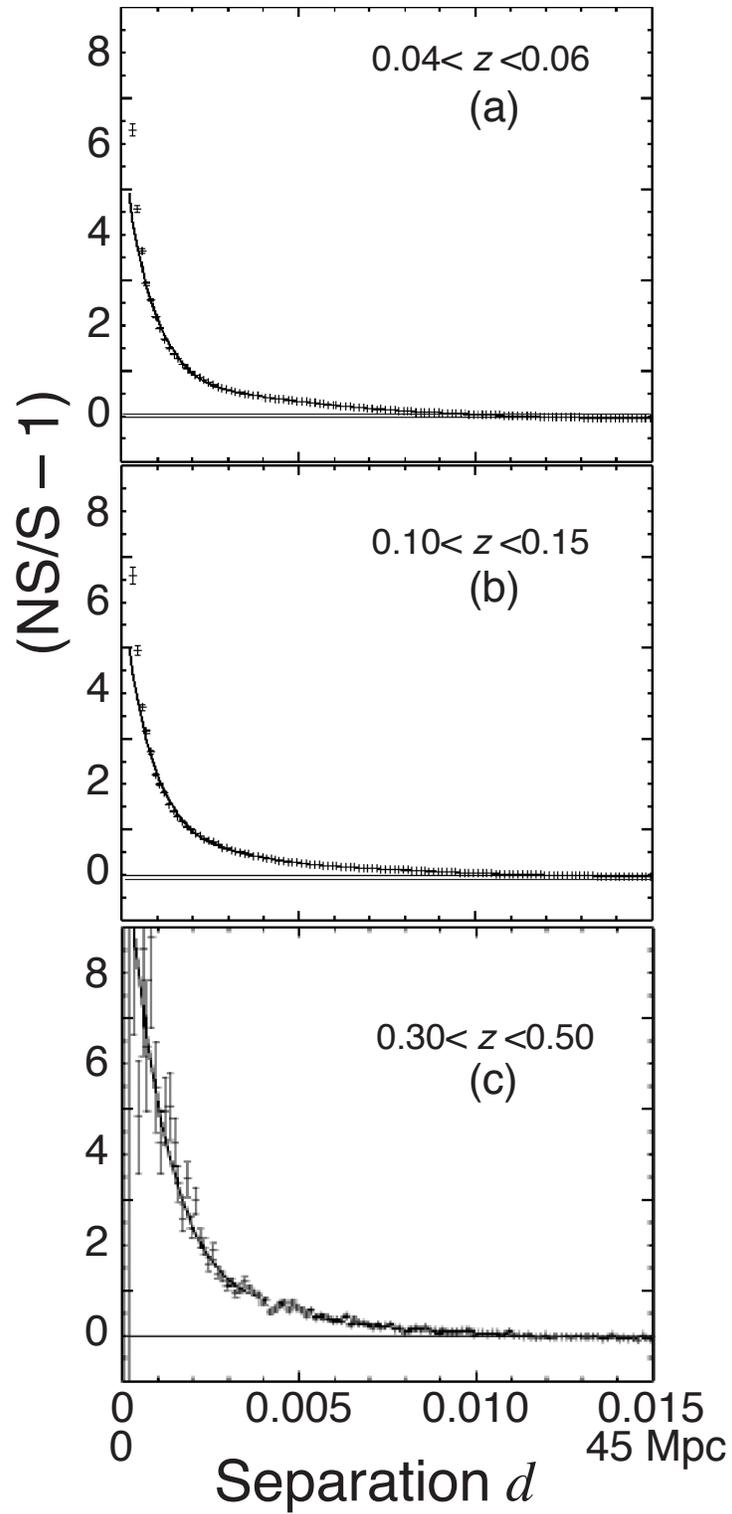

Figure 1 – Typical plots of *NotScrambled/Scrambled* correlation *vs*. separation in dimensionless units and Mpc for 3 redshift ranges. The smooth curves are fits to an exponential.

The effects of systematic uncertainties, such as edge effects, depend on the context. Galaxies at the edge of the defined volumes will have neighbors on one side only. These effects are expected to be small as long as the dimensions of the slice are much larger than the cluster size ~9 Mpc/$h$. [For example, the volumes for $0.2 < z < 0.3$ were ~$(500$ Mpc/$h)^3$.] Systematics due to edge effects were minimized by using the same size angle bins when looking for possible variations in the radii and integrals with right ascension and declination. Tests for this were made by dividing the slices into smaller ranges of right ascension and declination. No significant changes in the parameters were seen for $z > 0.04$. Regions near the limits of the SDSS coverage were also not used. For a given redshift the angle bins all had the same volume and the numbers of galaxies in each were approximately the same.

The redshift dependence is more complicated. The spatial density of observed galaxies decreases with increasing $z$ since only the brightest galaxies are above the detection threshold at higher $z$. Therefore, at larger $z$ the composition of the SDSS sample changes to include only the brightest, most massive galaxies that are more strongly clustered. Higher $z$ also corresponds to earlier epochs when the clusters were more diffuse (*i.e.*, larger comoving radii and lower amplitudes or integrals). The question of the $z$ dependence has been discussed in many references and will not be addressed here. See, for example, Ref. [16, 18, 19] and references cited there.

### 4. VARIATION OF RADII AND INTEGRALS WITH $\alpha$, $\delta$ AND GALAXY DENSITY

Figure 2 compares the correlation radii and integrals for 5 right ascension bins, each 24° wide for 4 redshift ranges. Declinations between 0° and 60° were used. Error bars that are smaller than the symbol sizes are not shown. As discussed in Sec. 6, the observed variations of $r_{sum}$ and $I_{sum}$ with $\alpha$ are consistent with that expected from cosmic variance. The $I_{sum}$ show a significant evolution with increasing $z$ due to the selection effects in the SDSS sample. Similar plots (not shown) of these quantities *vs.* $\delta$ show no significant variation with $\delta$ in the range 0° to 60°.

The variation of $r_{sum}$ and $I_{sum}$ with galaxy number density could be studied by choosing a fraction of the SDSS galaxies at random. For example, with ½ of the galaxies, $r_{sum}$ increased by 17% and $I_{sum}$ decreased by 31% for $0.20 < z < 0.30$. In the analysis, for a given $z$ range the volume of each right ascension bin was kept constant and the number of galaxies in each bin was the same to within about 8% for the higher redshift ranges and 14% for $0.08 < z < 0.10$. The variations of galaxy counts in the different $z$ and $\alpha$ bins did not appear to be correlated with $\alpha$.

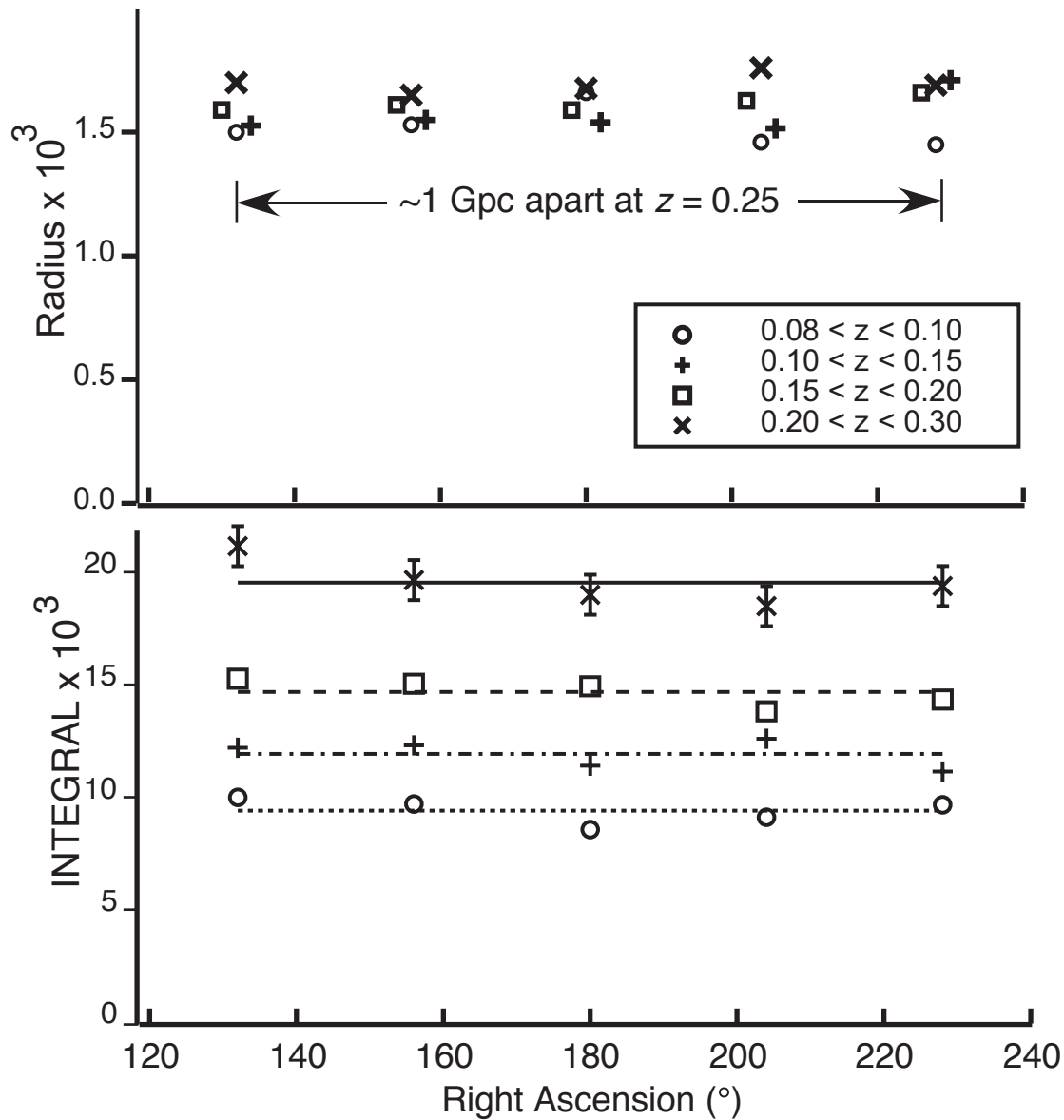

Figure 2 – Correlation radii and integrals *vs*. right ascension $\alpha$ for 4 redshift ranges. Except where shown, the uncertainties are smaller than the symbols. The integrals show a significant evolution with increasing redshift. The horizontal lines show the average integrals for 4 redshift bins. Results for $0.3 < z < 0.5$ and $z < 0.08$ have larger statistical errors and are not shown.

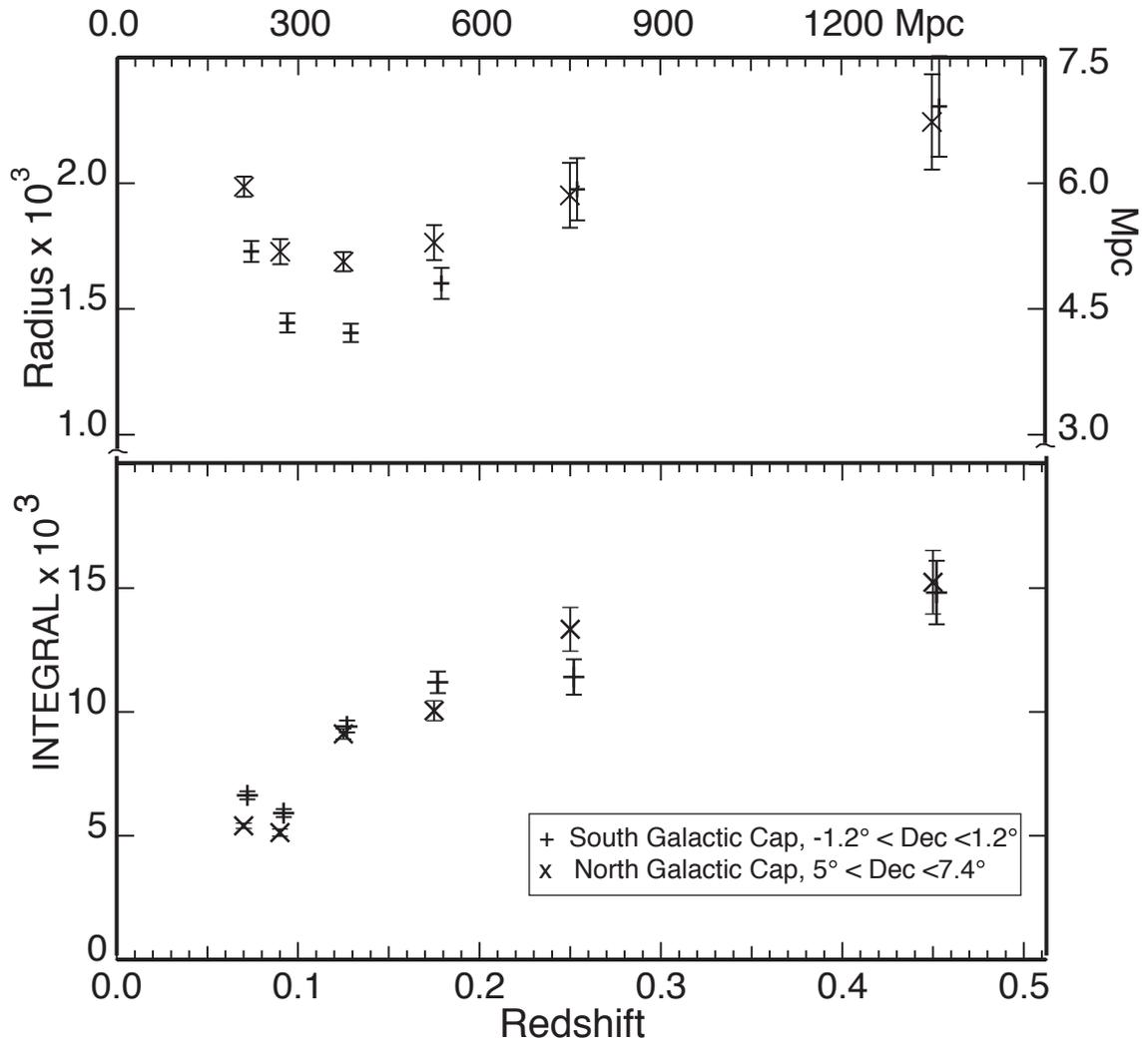

Figure 3 – Comparison of correlation radii and integrals at different redshifts for regions in the NGC and SGC with similar geometry and spatial density. For $z \sim 0.4$, these regions are separated by ~2.4 Gpc. The scale on top gives the separation in Mpc/$h$. The observed differences between the NGC and SGC at the same redshift are generally consistent with variations expected from cosmic variance for the small volumes studied.

## 5. COMPARISON OF THE NORTH AND SOUTH GALACTIC CAP REGIONS

It is of interest to compare the correlation parameters over regions of space as far apart as possible. Most of the SDSS galaxies with spectra are in the north Galactic cap region toward ($\alpha$, $\delta$) = (193°, 27°). For the hemisphere toward $\alpha = 0°$, only narrow bands in $\delta$ near −10°, 0°, and +10° are covered. This disparity complicates a comparison of the NGC and SGC regions. To allow this comparison, two slices with a width of 2.4° in $\delta$ and in opposite hemispheres were selected. Both slices included a range of about 100° in $\alpha$. When necessary, the appropriate numbers of galaxies in the NGC were randomly selected so that the spatial densities of galaxies in both hemispheres were approximately equal. The $r_{sum}$ and $I_{sum}$ for 6 ranges of redshift are compared in Fig. 3. At the higher redshifts the $r_{sum}$ and $I_{sum}$ at the same $z$ are in good agreement for the two hemispheres. However, for $z \lesssim 0.2$ there is some evidence of larger $r_{sum}$ and slightly smaller $I_{sum}$ toward the NGC, but it is likely that this small discrepancy is due to edge effects. The regions studied are only 2.4° wide in declination and the volumes are quite small, especially at the lower redshifts. Despite the attempt to match the geometries and spatial density in the two regions it is not possible to exactly match the spatial distributions within the slices. In any case, the volumes of the slices involved are $\lesssim 0.01$ Gpc$^3$. Fluctuations much greater than 10% are expected for volumes that small within the standard model [25], as discussed further in the next section. On the other hand, the good agreement for the larger volumes with $z > 0.2$ shows that the correlation radii and amplitudes are very similar for regions of space at equal $z$ that are separated by $\approx 2.4$ Gpc from each other.

## 6. DISCUSSION

There have been surprisingly few explicit tests of cosmic homogeneity and isotropy, despite its fundamental importance in cosmology. In fact, the latest Planck results seem to show a "hemispherical asymmetry" with a probability <0.008 at smoothing scales of 5° [13]. The hemispherical asymmetry has an axis roughly at right ascension 108° with $\delta \approx -10°$. A similar anisotropy was seen in the WMAP data. (See, for example, [26, 27].) This presumably is due to an over-density toward the northern ecliptic hemisphere. The data in Fig. 2 span $\alpha = 130° - 230°$. Unfortunately, the SDSS data do not extend to 108°, which is very close to the Galactic disk, so it is difficult to check for a locally over-dense region there using galaxy survey data. The $I_{sum}$ data for $0.20 < z < 0.30$ in Fig. 2 do hint at an increase at $\alpha \lesssim 150°$.

Table I gives the mean $I_{sum}$ and $r_{sum}$ for the 5 equal-volume $\alpha$ slices in Fig. 2 at each redshift range and their standard deviations calculated from the spread among the 5 slices. Mean galaxy counts, $n_g$, in each are also given. Watson et al. [25] estimate the fluctuations in galaxy counts ~10% due to cosmic variance would be expected in the $\Lambda$CDM model for volumes ~(0.5

| Redshift range | Mean $I_{sum}$ | Mean $r_{sum}$ | Mean $n_g$ | Volume (Gpc$^3$) | Expected variation |
|---|---|---|---|---|---|
| 0.30 – 0.50 | 23.10 ± 3.5% | 1.96 ± 2.0% | 12499 ± 3.1% | 21.1 | 1.5% |
| 0.20 – 0.30 | 19.54 ± 5.1% | 1.695 ± 2.4% | 9124 ± 4.9% | 4.9 | 3.2% |
| 0.15 – 0.20 | 14.68 ± 4.0% | 1.615 ± 1.9% | 16102 ± 6.2% | 1.3 | 6.2% |
| 0.10 – 0.15 | 11.93 ± 5.2% | 1.568 ± 5.1% | 31179 ±7.2% | 0.69 | 8.5% |
| 0.08 – 0.10 | 9.41 ± 6.0% | 1.520 ± 5.6% | 14771 ± 11.1% | 0.15 | 18.4% |

Table I – Parameters for fit of Integrals and Radii in Fig. 2 to a constant for the 5 equal-volume right ascension slices. The standard deviations for $I_{sum}$ and $r_{sum}$ are calculated from the variation among the 5 slices and are given in percent. The $n_g$ are the galaxy counts. The approximate comoving volumes for each slice are shown in the next-to-last column. The last column gives the fluctuations expected from cosmic variance based on $\Lambda$CDM estimates in Ref. [25].

Gpc/$h$)$^3$. The last column in Table I gives the expected variation in these parameters based on the Watson et al. estimate with the assumption that the variation scales as (Volume)$^{-\frac{1}{2}}$. The observed standard deviations for the $n_g$ and $I_{sum}$ are consistent with that expectation. The Watson *et al*. simulations [25] do not include smearing of the cluster radii due to the relative motion of galaxies within the cluster and other important details. They therefore cannot make reliable predictions for the variations in $r_{sum}$ for the right ascension slices.

The correlation lengths found here, ≈4.8 Mpc/$h$, are roughly consistent with other measurements (Zehavi *et al*. [28]; Davis & Peebles [29]), though the radii are definition dependent. Zehavi *et al*., using the standard $\xi(r) = (r/r_0)^{-\gamma}$ fit to SDSS data, find that $r_0$ increases from 2.83±0.19 Mpc/$h$ for R-band luminosity $M_r \approx -17.5$ and $z \approx 0.02$ to 10.00±0.29 Mpc/$h$ for $M_r \approx -22.5$ and $z \approx 0.17$. Carlberg *et al*. [30], analyzing the CNOC2 high-luminosity sample with a luminosity-compensated absolute magnitude, find that $r_0$ decreases from 4.75±0.05 Mpc/$h$ for $z \approx 0.10$ to 4.26±0.18 Mpc/$h$ for $z \approx 0.49$. None of the previous analyses attempt to compare the correlation amplitudes and none compare different regions of the sky at the same redshift.

There has been considerable discussion of alternatives to dark energy that attempt to explain the cosmic microwave background data and the Type Ia supernovae data using an inhomogeneous universe. One alternative model supposes that we are located near the center of a large, underdense, nearly spherical, void. (See, for example, Clifton, Ferreira, and Land [31]; Zibin, Moss & Scott [32].) While this analysis does not address the radial dependence question, it does show that we would have to be within several hundred Mpc/$h$ from its center.

## 7. CONCLUSIONS

The concordance $\Lambda$CDM model of the Universe assumes the cosmological principle that the Universe is homogeneous and isotropic on sufficiently large scales. The cluster radii and amplitudes from the SDSS data provide a direct test for inhomogeneity. The data in Fig. 2 and Table I show that the cluster radii and amplitudes are constant to about 5% over distance scales ~1 Gpc. The comparison of the north Galactic cap region with the south cap in Fig. 3 shows that they are equal to about 10% for regions separated by ~2.4 Gpc in the other direction. These findings are consistent with expectations of cosmic variance in the $\Lambda$CDM model [25].

The correlation parameters at constant $z$ appear to be consistent in all directions, so that the void model that poses an alternative explanation to the dark energy paradigm is only tenable if we happen to be near the center of a large void that has a galaxy density ~$1/4^{th}$ that of the surrounding universe.

This analysis would not have been possible without the dedicated efforts of the SDSS collaboration.

REFERENCES

[1]   L. Dai, D. Jeong, M. Kamionkowski, and J. Chluba, "The pesky power asymmetry", Phys. Rev. D **87,** 123005 (2013).

[2]   K. Land and J. Magueijo, **"**Examination of Evidence for a Preferred Axis in the Cosmic Radiation Anisotropy**"**, Phys. Rev. Lett. 95, 071301 (2005).

[3]   A. Gruppuso, P. Natoli, F. Paci, F. Finelli, D. Molinari, D., A. De Rosa, and N. Mandolesi, "Low variance at large scales of WMAP 9 year data", JCAP **07**, 047 (2013).

[4]   D. Schwarz, G. D. Starkman, D. Huterer, and C. J. Copi, "Is the Low-ℓ Microwave Background Cosmic? ", Phys. Rev. Lett. **93,** 221301 (2004).

[5]   A. Moss, D. Scott, J. P. Zibin, and R. Battye, "Tilted physics: A cosmologically dipole-modulated sky", Phys. Rev. D **84,** 023014 (2011).

[6]   F. R. Urban and A. R. Zhitnitsky, "P-odd universe, dark energy, and QCD", Phys. Rev. D **83,** 123532 (2011).

[7]   M. Hansen A. M. Frejsel, J. Kim, P. Naselsky, and F. Nesti, "Pearson's random walk in the space of the CMB phases: Evidence for parity asymmetry", Phys. Rev. D **83,** 103508 (2011).

[8]   A. Mariano and L. Perivolaropoulos, "CMB maximum temperature asymmetry axis: Alignment with other cosmic asymmetries", Phys. Rev. D **87,** 043511 (2013).

[9]   H. A. Feldman, R. Watkins, and M. J. Hudson, "Cosmic flows on 100 $h^{-1}$ Mpc scales: standardized minimum variance bulk flow, shear and octupole moments", Mon. Not. R. Astron. Soc. **407,** 2017 (2010).

[10]   J. K. Webb, J. A. King, M. T. Murphy, V. V. Flambaum, R. F. Carswell, and M. B. Bainbridge, "Indications of a Spatial Variation of the Fine Structure Constant", Phys. Rev. Lett. **107**, 191101 (2011).

[11]   A. Mariano and L. Perivolaropoulos, "Is there correlation between fine structure and dark energy cosmic dipoles? ", Phys. Rev. D **86**, 083517 (2012).

[12]   C. Bennett *et al*., "Seven-year Wilkinson Microwave Anisotropy Probe (WMAP) Observations: Are There Cosmic Microwave Background Anomalies? ", Astrophys. J. Suppl. **192**, 17 (2011).

[13] P. Ade *et al.* (Planck Collaboration), arXiv:1303.5083v1, "Planck 2013 results XXIII. Isotropy and statistics of the CMB", [to be published in Astron. Astrophys].

[14] A. R. Liddle and M. Cortês, "Cosmic Microwave Background Anomalies in an Open Universe", Phys. Rev. Lett. **111**, 111302 (2013).

[15] M. I. Scrimgeour *et al.*, "The WiggleZ Dark Energy Survey: the transition to large-scale cosmic homogeneity", Mon. Not. R. Astron. Soc. **425**, 116 (2012).

[16] H. Guo *et al.*, "The Clustering of Galaxies in the SDSS-III Baryon Oscillation Spectroscopic Survey: Luminosity and Color Dependence and Redshift Evolution", Astrophys. J., **767**, 122 (2013).

[17] A. R. Pullen and C. M. Hirata, "Systematic Effects in Large-Scale Angular Power Spectra of Photometric Quasars and Implications for Constraining Primordial Non-Gaussianity", PASP **125**, 705 (2013).

[18] D. Parkinson *et al.*, "The WiggleZ Dark Energy Survey: Final data release and cosmological results", Phys. Rev. D **86**, 103518 (2012).

[19] A. J. Ross *et al.*, "The Clustering of Galaxies in the SDSS-III DR10 Baryon Oscillation Spectroscopic Survey: No Detectable Colour Dependence of Distance Scale or Growth Rate Measurements", Mon. Not. R. Astron. Soc. **424**, 564 (2012).

[20] P. Ade *et al.* (Planck Collaboration), arXiv:1303.5080v1, "Planck 2013 results XX. Cosmology from Sunyaev-Zeldovich cluster counts", [to be published in Astron. Astrophys].

[21] For a discussion of biases in cluster detection, see M.D. Gladders, H. K. C. Yee, "A New Method For Galaxy Cluster Detection. I. The Algorithm", Astrophys. J. **120**, 2148 (2000) and references cited there.

[22] J. R. Whitbourn and T. Shanks, "The local hole revealed by galaxy counts and redshifts", Mon. Not. R. Astron. Soc. **437**, (2014)

[23] J. R. Gott, M. Jurić, D. Schlegel, F. Hoyle, M. Vogeley, M. Tegmark, N. Bahcall, and J. Brinkmann, "A Map of the Universe", Astrophys. J. **624**, 463 (2005).

[24] K. Abazajian *et al.*, "The Seventh Data Release of the Sloan Digital Sky Survey", Astrophys. J. Suppl. **182,** 543 (2009).

[25] W. A. Watson, , I. T. Iliev, J. M. Diego, S. Gottlöber, A. Knebe, E. Martínez-González, G. Yepes, "Statistics of extreme objects in the Juropa Hubble Volume simulation", Mon. Not. R. Astron. Soc. **437**, 3776 (2014).

[26]  H. Eriksen, F. K. Hansen, A. J. Banday, K. M. Górski, K., P, B. Lilje, "Asymmetries in the Cosmic Microwave Background Anisotropy Field", Astrophys. J. **605,** 14 (2004).

[27]  M. Axelsson, Y. Fantaye, F. K. Hansen, A. J. Banday, H. K. Eriksen, and K. M. Gorski, "Directional Dependence of ΛCDM Cosmological Parameters", Astrophys. J. **773**, L3 (2013).

[28]  I. Zehavi *et al*., "The Luminosity and Color Dependence of the Galaxy Correlation Function", Astrophys. J. **630**, 1 (2005).

[29]  M. Davis, P.J. Peebles, "A survey of galaxy redshifts. V - The two-point position and velocity correlations", Astrophys. J. **267**, 465 (1983).

[30]  R. G. Carlberg, H. K. C. Yee, S. L. Morris, H. Lin, P. B. Hall, D. Patton, M. Sawicki, C. W. Shepherd, "Galaxy Clustering Evolution in the CNOC2 High-Luminosity Sample", Astrophys. J. **542,** 57 (2000).

[31]  T. Clifton, P. Ferreira, K. Land, "Living in a Void: Testing the Copernican Principle with Distant Supernovae", Phys. Rev. Lett. **101**, 1313021 (2008).

[32]  J. P. Zibin, A. Moss, and D. Scott, "Can We Avoid Dark Energy? ", Phys. Rev. Lett. **101** 251303 (2008).